\documentclass[twocolumn,tighten]{aastex62}

\usepackage{placeins}
\usepackage{natbib}
\usepackage{amsmath}
\usepackage{color}
\usepackage{enumerate}
\usepackage{graphicx}
\usepackage{soul}

\shorttitle{Magnetic Protostellar Cores}
\shortauthors{Kuznetsova et al.}

\begin{document}


\title{Angular Momenta, Magnetization, and Accretion of Protostellar Cores }

\correspondingauthor{Aleksandra Kuznetsova}
\email{kuza@umich.edu}

\author{Aleksandra Kuznetsova}
\affil{Department of Astronomy, University of Michigan, 1085 S. University Ave., Ann Arbor, MI 48109}
\author{ Lee Hartmann}
\affil{Department of Astronomy, University of Michigan, 1085 S. University Ave., Ann Arbor, MI 48109}
\author{Fabian Heitsch}
\affil{Department of Physics and Astronomy, University of North Carolina - Chapel Hill}


\begin{abstract}

Building on our previous hydrodynamic study of the angular momenta of cloud cores formed during gravitational collapse of star-forming molecular gas in \cite{kuznetsova2019}, we now examine core properties assuming
ideal magnetohydrodynamics (MHD). Using the same sink-patch implementation for the \emph{Athena} MHD code, we characterize the statistical properties of cores, including the mass accretion rates, specific angular momenta, and alignments between the magnetic field and the spin axis of the core on the $0.1 \ \mathrm{pc}$ scale. 
Our simulations, which reproduce the observed relation between magnetic field strength and gas density, show that magnetic fields can help collimate low density flows and help seed the locations of filamentary structures.
Consistent with our previous purely hydrodynamic simulations, stars (sinks) form within the heterogeneous environments of filaments, such that accretion onto cores is highly episodic leading to short-term variability but no long-term monotonic growth of the specific angular momenta. With statistical characterization of protostellar cores properties and behaviors, we aim to provide a starting point for building more realistic and self-consistent disk formation models, helping to address whether magnetic fields can prevent the development of (large) circumstellar disks in the ideal MHD limit.
\end{abstract}

\keywords{stars:formation, stars: kinematics and dynamics, ISM: kinematics and dynamics}

\citestyle{aa}

\section{Introduction}

While magnetic fields in the interstellar medium are clearly dynamically important at low densities \citep[see review by][]{crutcher2012}, their importance in high-density regions of molecular clouds, and thus on the formation of stars and circumstellar disks,
is less clear.
Observations on scales of $\sim 0.1 - 10$~pc \citep[e.g.,]{crutcher2012,palmeirim2013,planck2016}
show that magnetic fields are important in forming filaments, as suggested by preferential
orientation of filaments perpendicular to fields.

In addition, recent calculations of protostellar core collapse have proposed that magnetic fields could restrain or even prevent
protoplanetary
disk formation by transporting large
amounts of angular momentum outward-
the so-called
``magnetic braking catastrophe” \citep{mellon2008}.  

\par
Exactly how effective magnetic fields are in suppressing disk formation depends upon a variety of parameters, in addition to the possible importance of non-ideal magnetohydrodynamic (MHD) effects.
The simulations
suggest that the amount of magnetization $\mu$ - the mass-to-flux ratio ($M/\Phi$) over the critical value, $(M/\Phi)_{crit}  = 1/2\pi\sqrt{G}$ \citep[for a sheet; ][]{nakano1978}
- and the geometry of the magnetic field, especially relative to the angular momentum vector, determine the frequency, shape, and size of protoplanetary disks that can form \citep{joos2012,li2013,krumholz2013}. 
It seems that without the dissipation of magnetic flux, whether through ambipolar diffusion \citep{mellon2008,vaytet2018}, or other non-ideal effects such as the Hall effect \citep{wurster2018}, a protostellar core with an initially strong magnetic field that is aligned with the primary rotation axis may not be able to form disks having sizes large enough to explain observations \citep[e.g.,][]{Tobin_2012,ohashi14}.
Turbulence in the protostellar core as a cause of misalignment \citep{joos2013, gray2017}, reconnection diffusion of the magnetic field \citep{santos2013}, or as a method to halt the production of large toroidal components responsible for braking \citep{seifried2012} has also been previously suggested as a means of reconciling magnetized disk formation. 

\par The median magnetization parameter of cores derived from Zeeman effect measurements in the interstellar medium is estimated to be $\mu \sim ~2$ \citep{crutcher2012}, indicating that fields are dynamically important.  Observational constraints on the relative directions of magnetic fields and angular momenta, assuming that the latter are given by the positions of bipolar outflows, suggest random alignment, at least on 1000 AU scales \citep[see review by][]{hull2019}.However, \citet{krumholz2013} contended that at that magnetization even with a random distribution of angles between the magnetic field and angular momentum vector, the disk frequency would still be limited to ~10-15\%, much lower than the observed Class II disk fraction. 

\par Top-down cluster formation simulations, starting with the progenitor molecular cloud down to protostellar core and disk scales, can connect the conditions of star forming environments to disk formation scenarios \citep{kuffmeier2017,Bate_2018} and produce statistical descriptions of protostellar properties and behaviors \citep{Kuznetsova_2018b,kuznetsova2019}. Top-down simulations can also characterize the multi-scale behavior of magnetic fields during star formation. Connecting large scale behavior with properties on the core scale can provide realistic initial conditions for core collapse and disk formation studies \citep[e.g.][]{bhandare2018}
, given the inherent ambiguities between scattering effects and the magnetic field geometry seen in disk scale polarization studies \citep{kataoka2015,yang2016}.

\par In this paper, we present results of ideal magnetohydrodynamic (MHD) simulations performed with the grid code \emph{Athena}. Large scale magnetic field behavior follows the
weak-field model in which magnetic fields play a role in assembling material on large scales and statistical analysis of filament and magnetic field orientation is in agreement with findings from observed young star forming regions \citep{planck2016}. We find that the presence of magnetic fields slows star formation compared to the purely hydrodynamic case, but given the weak initial field, gravity remains the primary driver of star formation leading to the formation of supercritical cores. On core scales, we find similar mass and angular momentum accretion behavior to our hydrodynamic simulations in \citet{Kuznetsova_2018b,kuznetsova2019}. As in \citet{kuznetsova2019} (KHH19), we find that the accretion of angular momentum is episodic and highly variable over short time scales, but appears constant when time averaged. The addition of magnetic fields in this work was found to dampen the amplitude of variability in the specific angular momentum. 

\par Our simulations provide an important comparison to newly released work \citep{wurster2019}, in that we use a completely different numerical method (fixed grid in our case vs. SPH in \citep{wurster2019}), and the larger number of cores in our work provide a more complete statistical characterization of core angular momenta and magnetizations for use by simulations of protostellar core collapse
on smaller scales.

\section{Method}

\par We use a version of the Eulerian grid code \emph{Athena} \citep{Stone_2008}, modified with a sink-patch implementation described in \cite{kuznetsova2019} and an RK3 integrator \citep{GottliebShu1998}, to simulate star cluster formation from the progenitor molecular cloud. The basic numerical setup and details of the sink-patch implementation are described in  \citep{Kuznetsova_2018b, kuznetsova2019}, so we provide only a short summary here.

We solve the system of equations
\begin{eqnarray}
\label{eq:mass}
\frac{\partial\rho}{\partial t} + \nabla \cdot (\rho \mathbf{v}) & = & 0\\
\label{eq:mom}
\frac{\partial \rho \mathbf{v}}{\partial t} + \nabla \cdot (\rho\mathbf{vv} -\mathbf{BB} + P + \frac{\mathbf{B \cdot B}}{2}) & = & -\rho \nabla \Phi\\
\label{eq:induction}
\frac{\partial \mathbf{B}}{\partial t} - \nabla \times (\mathbf{v} \times \mathbf{B}) & = & 0\\
\label{eq:poisson}
\nabla \Phi & = & 4\pi G \rho
\end{eqnarray}
in the ideal MHD limit. The equation of state is isothermal such that $P=c_s^2 \rho$. The Poisson equation (eq.~\ref{eq:poisson}) is solved every RK3 substep, using the FFT solver that comes with version 4.2 of \emph{Athena}. 

\par The sink-patch, which acts as the accretion reservoir for the sink particle, is defined by its accretion radius $r_{acc}$ such that it contains $(2r_{acc}+1)^3$ cells, centered on the location of the sink particle, redrawn at every timestep. Derived quantities are calculated on the boundary of the sink-patch.

\par Conserved variables were output for the patch cells and an additional boundary cell. Angular momentum is calculated as a cumulative sum over the patch cells, $\mathbf{L} = \Sigma_{i} m_i \mathbf{v}_i \times \mathbf{r}_i$ , where the velocity $\mathbf{v}_i$ and radius $\mathbf{r}_i$ are taken with respect to the sink's velocity and position, respectively. Specific angular momentum is then defined as $\mathbf{j} = \mathbf{L}/\Sigma_i m_i$. Fluxes of quantities into the sink-patch system were measured across the inner faces of the boundary cells, with information separable into orthogonal components such that the directional information of accretion onto the patch could be tracked. Sinks that have entered one another’s patches and merged, or those liable to fragment under the conditions outlined in KHH19 are removed from the analysis.

\par The initial conditions are that of a 4 pc radius spherical cloud in a 20 pc box at an isothermal temperature of $T = 14 \ \mathrm{K}$, the same as those used in \citet{Kuznetsova_2018b} and \citet{kuznetsova2019}, with the addition of a $5 \ \mu G$ uniform field.
 The exception to these conditions is run IR\_b0, which is purely hydrodynamic for comparison. We adopt an intermediate resolution (as designated in KHH19) of $N_{cell} = 512^3$ where $\Delta x = 0.04 \mathrm{pc}$, such that all of the runs in this paper have a patch radius of $r_p = (r_{acc} + 1/2)\Delta x = 0.1 \, \mathrm{pc}$. As the simulation has an isothermal equation of state, results can be rescaled to another family of solutions, preserving the sound and Alfven speed, by scaling temperature, distance, time, magnetic field strength, and mass by the same factor. 
 
 \begin{figure*}[ht]
\centering
\includegraphics[width=0.8\textwidth]{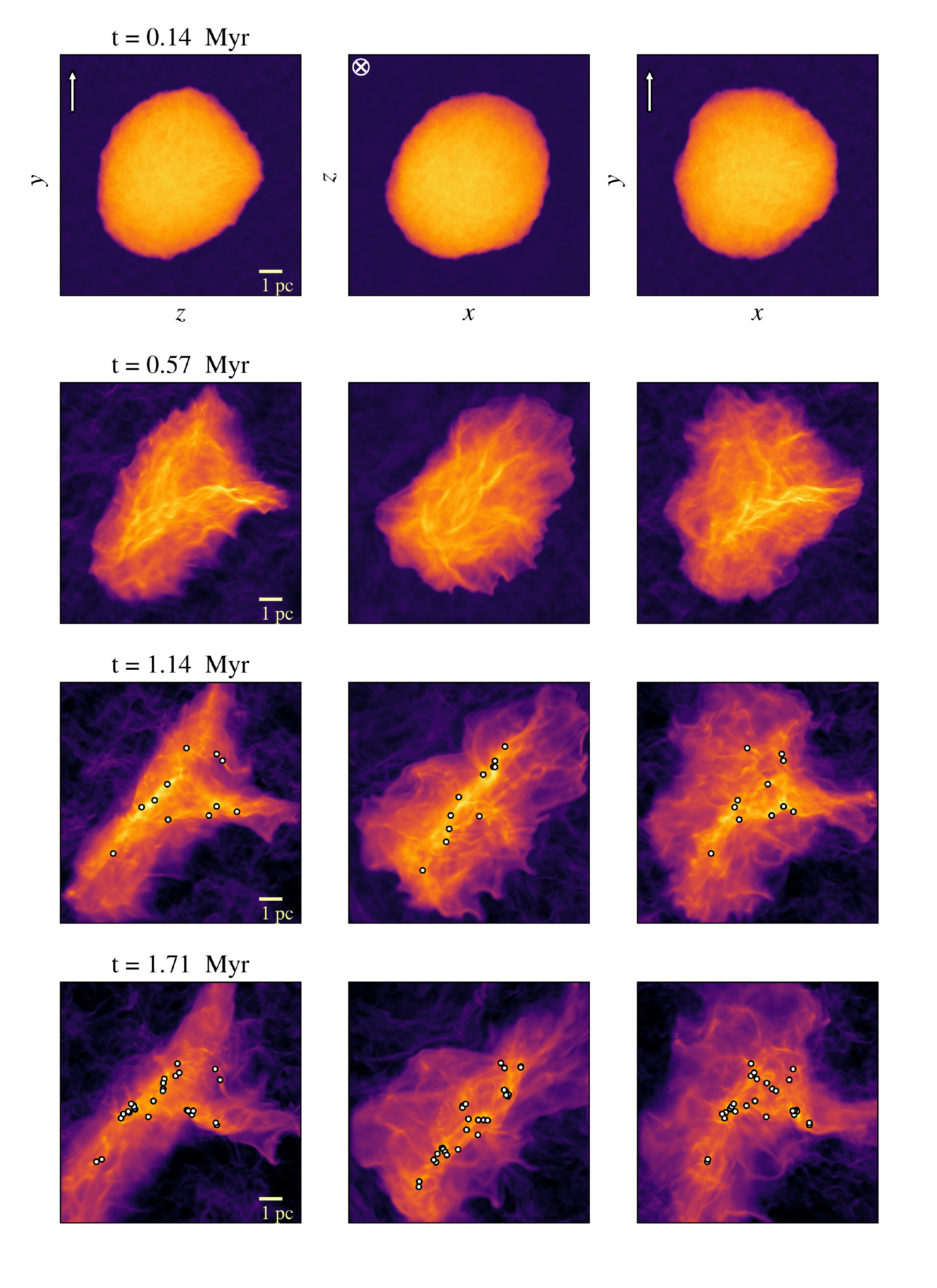}
\caption{Evolution of column density of run IR\_by5\_s2 through to one free-fall time, shown at t = 0.14, 0.57, 1.14, and 1.71 Myr at three orthogonal directions, labeled in the top-most row. Arrows and $\otimes$ shapes in the top-most row show the initial uniform magnetic field direction for each projection. Locations of sink particles are shown as white markers. Box shown is the inner $6 \mathrm{pc} \times 6 \mathrm{pc}$ of the domain, centered on the cloud; scale bar denotes 1 pc.  }
\label{fig:columnsinks}
\end{figure*}

 \par The ambient density exterior to the initial spherical cloud is $\rho_0 = 1.5 \times 10^{-23}$ \rm \ g \ $\mathrm{cm^{-3}}$, where $\rho_c = 100 \times \rho_0$, such that the initial cloud free-fall time is $t_\mathrm{ff} = 1.7$ Myr and initial cloud mass is $\sim 5900 \ \mathrm{M_{\odot}}$. Turbulence is seeded at the start with a Mach 8 velocity spectrum $P(k) \propto k^{-4} dk$ which introduces both overdensities and a base level of cloud scale angular momentum, $\Omega_k = 0.07 \pm 0.05 \ \rm km \ s^{-1} \ pc^{-1}$ or $j = 1.58 \times 10^{23} \ \rm cm^2 \  s^{-1}$ on average. Several turbulent realizations are used for the purpose of generating better core statistics. Details of all the runs are listed in Table \ref{tab:runs}. 

\begin{table}[h!]
\label{tab:runs}
\caption{List of run properties for the cores used in this study.}
\begin{tabular*}{\columnwidth}{lcccr}
\hline
Run         & Seed & $\Omega_k \ [\rm km \ s^{-1} \ pc^{-1}]$  & $\mathbf{B_0} \ \rm [\mu G]$                            & $N_\mathrm{{sink}}$ \\ \hline
IR\_b0      & 0    &    0.07                     & $\langle 0, 0, 0 \rangle$ &      108      \\ 
IR\_bx5     & 3    &    0.03                 & $\langle 5, 0, 0 \rangle $&        45    \\ 
IR\_bx5\_s1 & 1    &    0.13                    & $\langle 5, 0, 0 \rangle $&      17      \\ 
IR\_bx5\_s2 & 2    &    0.03                   & $\langle 5, 0, 0 \rangle$ &     9       \\ 
IR\_by5\_s1 & 1    &    0.13              & $\langle 0, 5, 0 \rangle$ &    15        \\ 
IR\_by5\_s2 & 2    &    0.03         & $\langle 0, 5, 0 \rangle$ &       35     \\ \hline
\end{tabular*}

\end{table}

\section{Results}

\par The initial perturbed cloud conditions are such that the virial parameter $\alpha_{vir} \lesssim 1$ on the cloud ($r\sim 4  \ \mathrm{pc}$) scale. As collapse proceeds, the onset of sheet/filament formation  creates dense structures that become the sites of later sink formation (Figure \ref{fig:columnsinks}). The simulations are evolved for up to one initial free-fall time ($t_{\mathrm{ff}}$ = 1.7 Myr).  Run IR\_by5\_s2 was evolved the furthest in time and produced the most sinks; we use it as the fiducial run for all cloud and filament scale analysis, unless otherwise stated. Statistical properties cores in subsection \ref{sect:cores} come from the full series of runs, and datasets are stacked where indicated.  

\par In Figure \ref{fig:columnsinks}, we show the progression of cluster formation in run IR\_by5\_s2 as a series of snapshots in time of column densities and sink locations, in three orthogonal projections. The first two columns in Figure \ref{fig:columnsinks} show the filament's long axis, while the line of sight of the  $x-y$ plane projection in the third column is preferentially along the elongated axis of the filament, that is, viewed more or less `down the barrel'. Thus, elongated features in the $z-y$ and $x-z$ planes are indicative of filament formation, but the appearance of non-spherical features and groupings in the last projection are more likely to be the result of projection effects, rather than co-spatiality.

\subsection{Global Magnetic Field Behavior}
\label{sec:globalb}
\begin{figure}[h!]
    \centering
    \includegraphics[width=0.95\columnwidth]{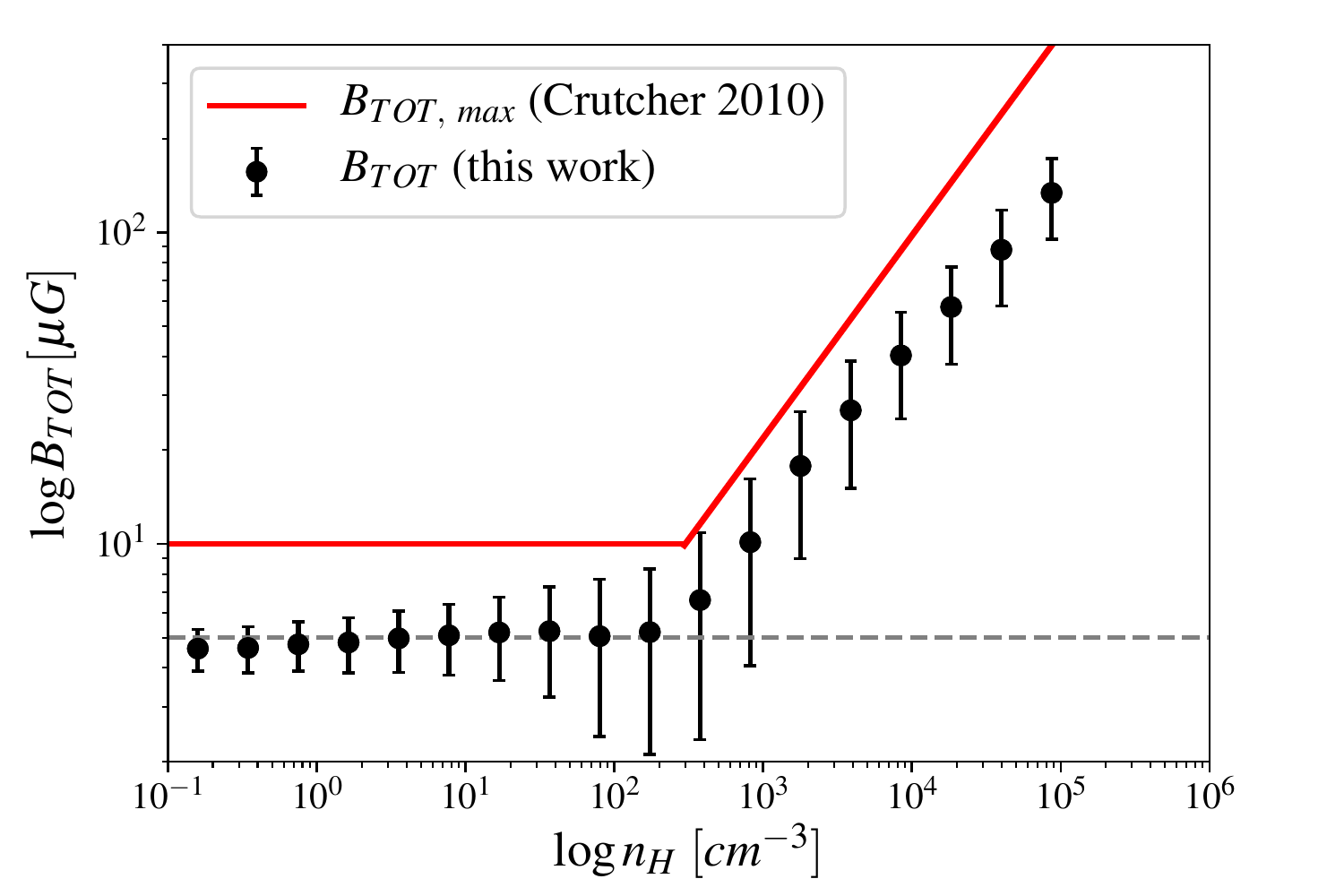}
    \caption{Median total magnetic field strength, $B_{TOT}$ within the respective binned number densities $n_H$. Binned cell averages are shown as black markers with error bars representing the standard deviation within the bin. The upper limit for the magnetic field, $B_{TOT,max}$, from \cite{crutcher2010} is shown in red. The gray dashed line is the initial magnitude of the magnetic field $B_0 = 5 \ \mu G$.}
    \label{fig:crutcher}
\end{figure}

An important test of our magnetic simulations  is whether they are consistent with the observed variation of magnetic field strength with gas density. Figure \ref{fig:crutcher} shows the results from the simulations, with number densities smoothed over 0.1 pc (consistent with our effective numerical resolution), along with the median total magnetic field in each density bin. The total magnetic field strength is largely independent of number density $n_H$, consistent with the initial magnitude of the magnetic field $B_0 = 5 \ \mu G$, until densities of $ n_{H} \sim 300-400 \ \mathrm{cm^{-3}}$, beyond which the median total magnetic field increases rapidly with increasing density.  The results are consistent with the maximum total magnetic field values as a function of density found by \cite{crutcher2010} from a Bayesian analysis of Zeeman observations of nearby clouds. The simulations are also consistent with the observational inference of a wide spread in magnetic field strengths at a given density \citep[see discussion in][]{crutcher2012}. We find a slope of median field strengths 
$B \sim n_H^{0.56}$ which is slightly flatter than the observationally estimated slope $0.65 \pm 0.05 $ for the maximum field strength
\cite{crutcher2012}, but this may be affected by limited resolution at the highest densities.

\begin{figure*}
    \centering
    \gridline{\fig{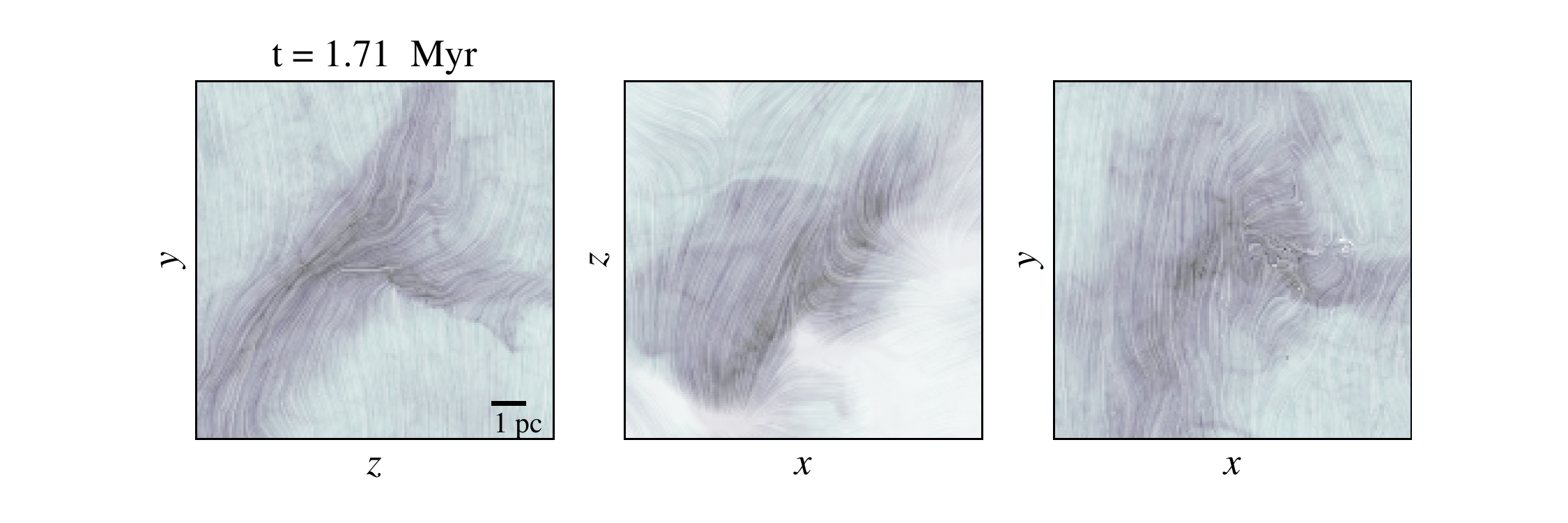}{0.85\textwidth}{}}
\gridline{\fig{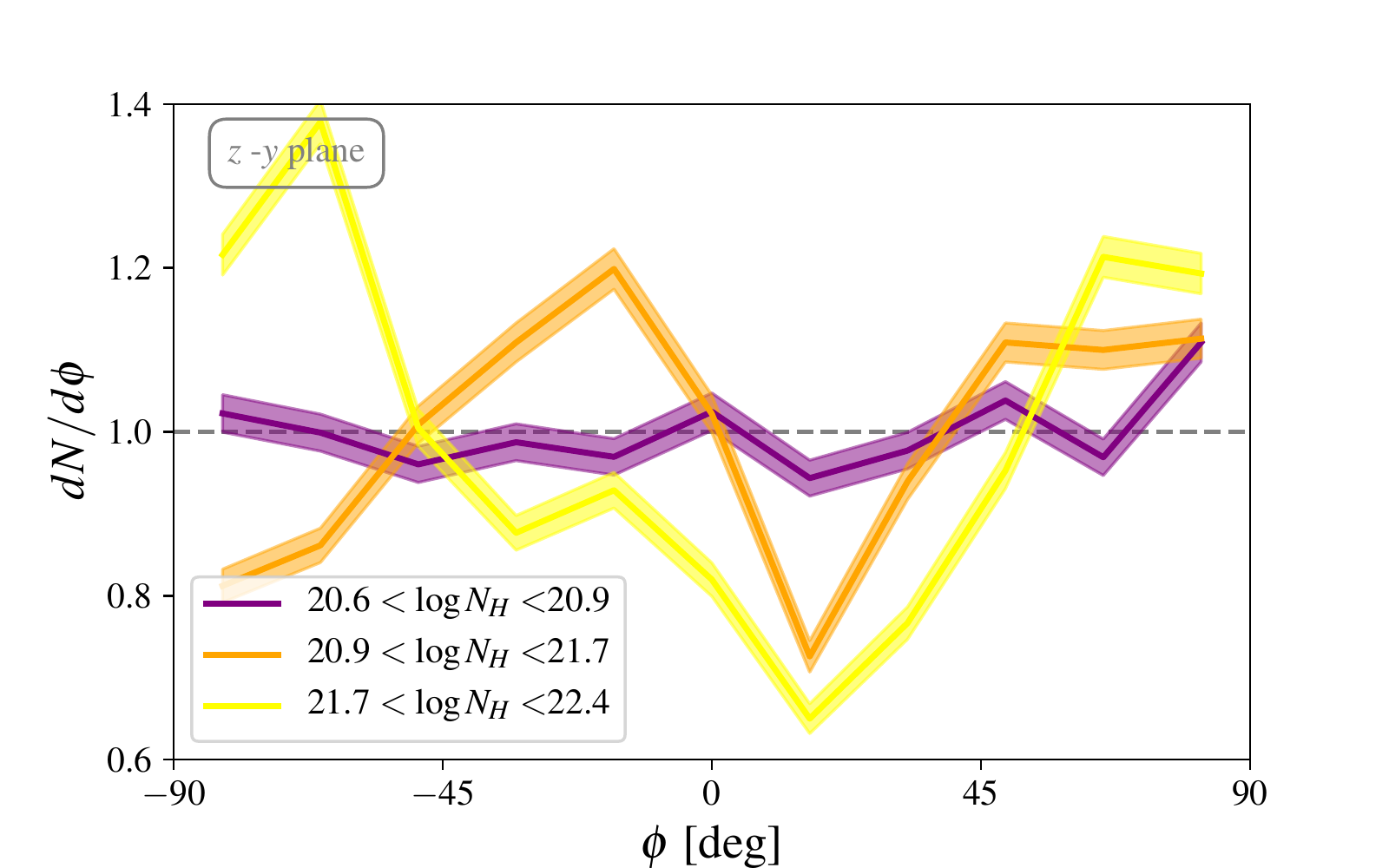}{0.35\textwidth}{(a)}
\fig{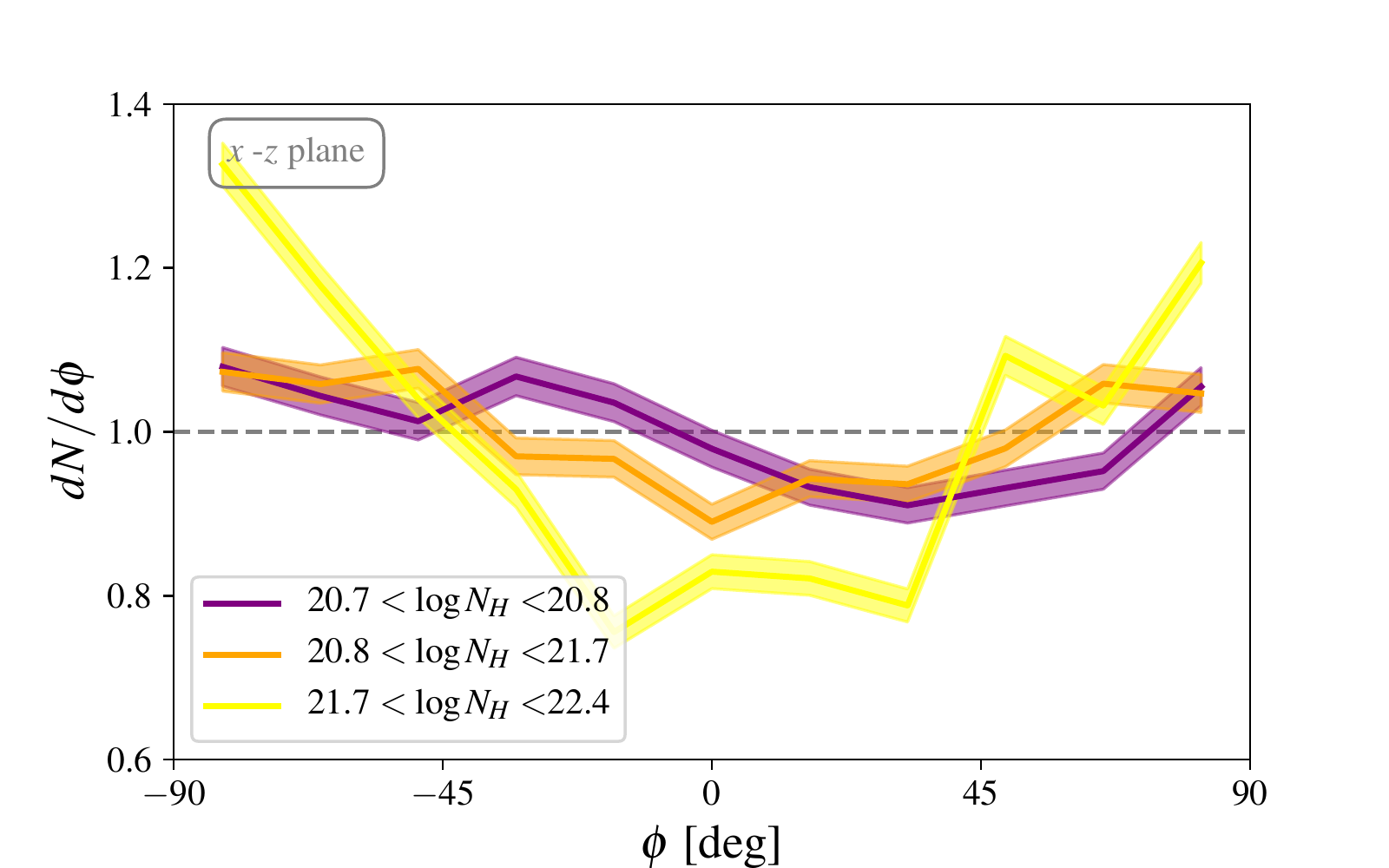}{0.35\textwidth}{(b)}
\fig{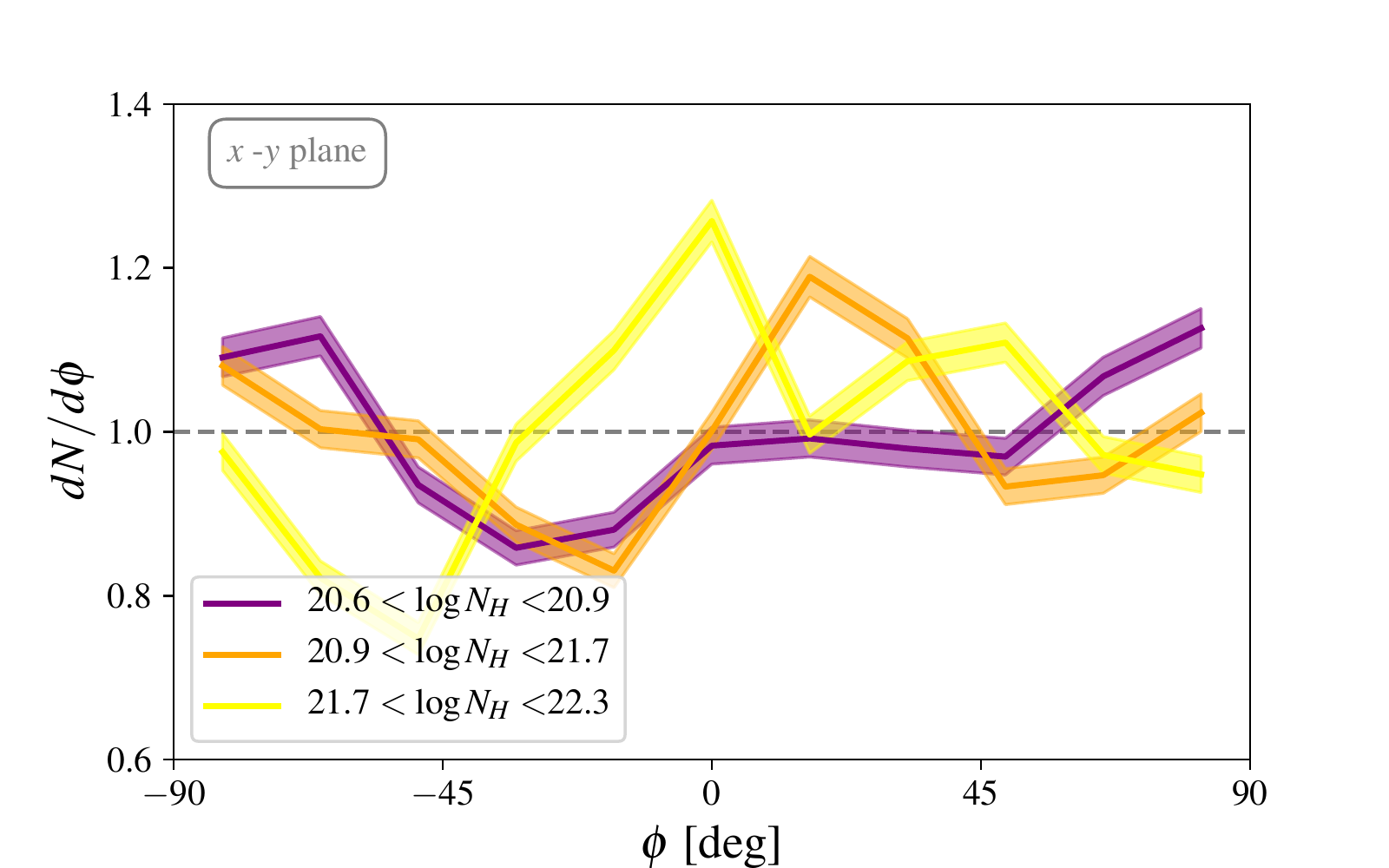}{0.35\textwidth}{(c)}}
    \caption{ (Top row) Magnetic field geometry computed as a line integral convolution (LIC), computed with the \texttt{LicPy} package, of the magnetic field shown on top of the column density in three projections at 1.71 Myr for run IR\_by5\_s2. Low-magnitude small scale noise in the LIC is smoothed and masked to appear as opaque regions in the map. The box sizes are $6 \ \rm pc \times 6 \rm \ pc$ of the simulation domain centered on the cloud. The initial magnetic field is uniform in the $y$-direction for this run. (Bottom row) Histograms of relative orientation (HRO) between the magnetic field direction and lines of isodensity computed using three different column density projections (smoothed to a resolution limit of $\sim 0.4 \ \rm pc$) in the (a) $z-y$ plane (b) $x-z$ plane (c) $x-y$ plane. Each line corresponds to the range of column densities at the 68th (purple), 90th (orange), and 98th (yellow) percentile level for each projection, with the same number of samples ($\sim 20000$), binned in increments of $15^{\circ}$. The error due to binning $\sigma_h$ is represented by the shaded contours for each line. The value of $dN/d\phi$ is normalized with respect to the expected amount of samples in each bin if the distribution of relative orientations were uniform.}
    \label{fig:HRO}
\end{figure*}

\par During global cloud collapse, magnetic fields aid in the collection of material into dense structures (filaments).  When filaments become massive enough, they begin to influence the magnetic field geometry; U-shaped bends around filaments as field lines get dragged in toward high density regions are particularly visible in the $y-z$ and $z-x$ projections at 1.71 Myr in Figure \ref{fig:HRO}. Smaller striation-like features tend to lie preferentially along the field lines, perpendicular to the main filament axis.

We computed
histograms of relative orientation (HROs) between the iso-column density contours and the plane of sky magnetic field in three different orthogonal projections (Figure \ref{fig:HRO}a-c). (Details for the calculation of the HROs based on the methods outlined in \citet{planck2016} are given in Appendix \ref{sect:hro}. The HROs show that the highest density structures (filaments) are preferentially perpendicular to large scale magnetic fields, consistent with {\em Planck} observations \citep[see Figures 3, 5, and 8 in]{planck2016}.  This suggests that regions in which there is no preferential anti-alignment detected in the HRO diagram could have a viewing geometry that misrepresents the degree of elongation or filamentary structure seen in column density maps. Thus, filaments are formed roughly perpendicular to the magnetic field, the original field orientation helping to seed the morphology of the final structure. In our simulations, this process works likely in combination with the initial turbulent field's largest scale fluctuations, which played a role in setting the filamentary morphology  of the purely hydrodynamic runs \citep{kuznetsova2019}.

\subsection{Properties on the Core Scale}

\begin{figure*}[htb]
    \centering
    \includegraphics[width=0.75\textwidth]{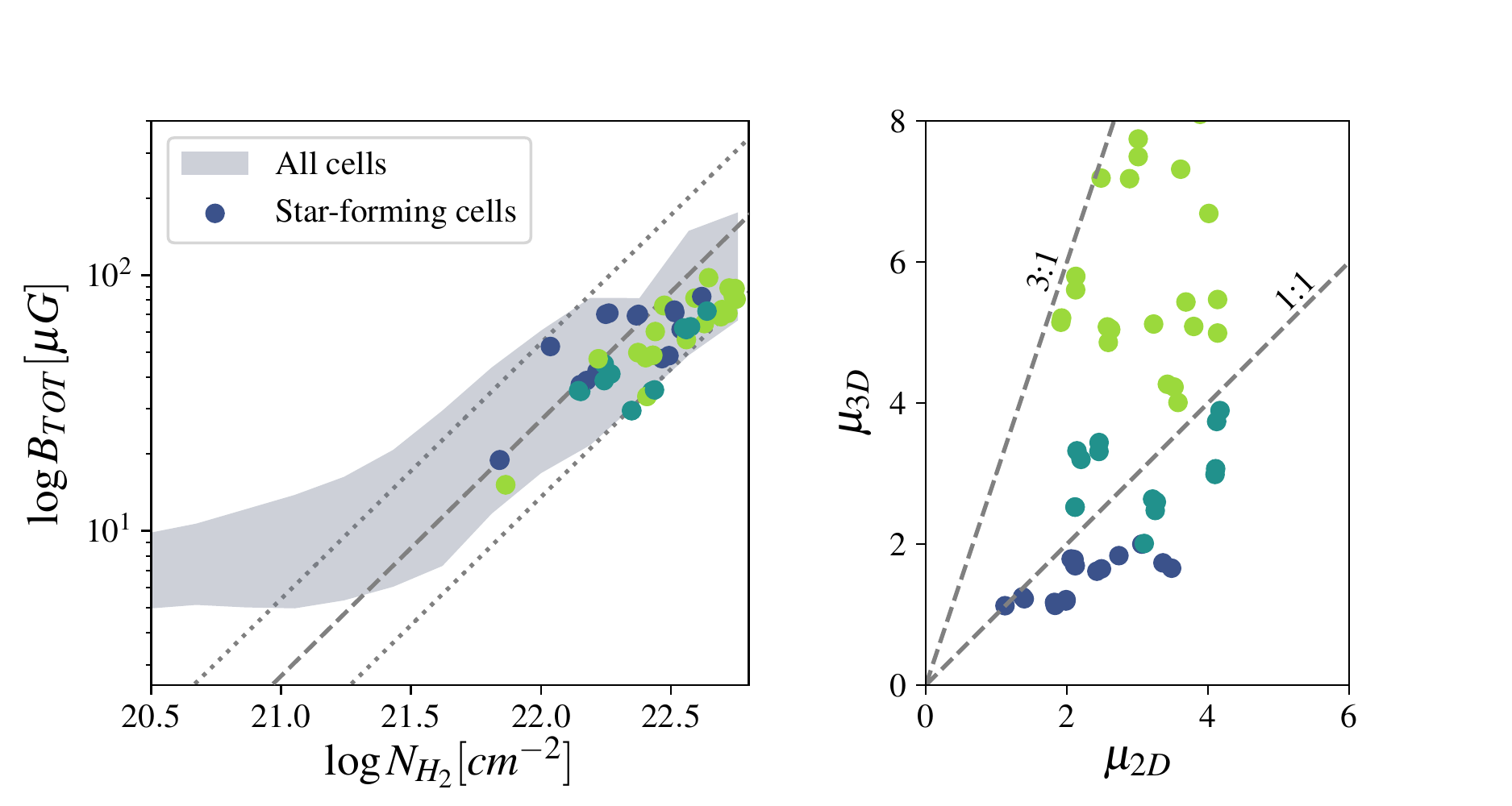}
    \caption{ (Left) Total magnetic field strength, $\log B_{TOT}$ vs column densities, $\log N_{H_2}$ of all cells in the cloud (gray region) and just the star forming cells -- those averaged across the sink-patch system (colored circular markers).  Lines of constant magnetization $\mu_{2D} = 1, 4$ are the upper and lower bounds indicated by the dotted line, with the $\mu_{2D} = 2$ dashed line in between, computed from \cite{crutcher2012}. Star-forming cells are binned by their true mass to flux ratios over the critical value $\mu_{3D}$, computed with the mass contained within the patch and the flux through the patch surface; colors correspond to values in the right plot: light green - $\mu_{3D} > 4$, teal - $2 < \mu_{3D} < 4 $, dark blue - $1 < \mu_{3D} < 2$, and purple - $\mu_{3D} < 1$.  
    Right: Comparison of the observed magnetization from a projection, $\mu_{2D}$, using relation from \cite{crutcher2012} vs the volumetrically derived magnetization, $\mu_{3D}$.}
    \label{fig:crutcherobs}
\end{figure*}

Individual sink-patch systems also exhibit the density and magnetic field scaling law seen in Figure \ref{fig:crutcher}.
In Figure \ref{fig:crutcherobs}a we show the relationship between the line of sight column density $N_{H_2}$ and the total magnetic field $B_{TOT}$  for a sample projection of the simulation cube, with colored circles denoting values taken at the locations of the sink-patches (``star-forming cells");  the star-forming cell values are computed as averages across the projected line of sight sink-patch cells and $B_{TOT}$ is a mass weighted average across the patch cells. (We use the patches as equivalent to molecular cloud cores-- see discussion in \cite{kuznetsova2019}). Each sink-patch system has a volumetrically- determined magnetization, $\mu_{3D}$, using the enclosed mass within the patch cells $M_p$ and the average flux through the patch surface as the dot product of the three averaged components of the patch with their areas $\Phi = \mathbf{B}\cdot dA$, indicated by their color.  For comparison we also compute an ``observational estimate'' of the critical ratio, using the relation $\mu_{2D} = N_{H_2} (7.6\times 10^{-21})/B_{TOT} /M/\Phi_{crit}$  \citep{crutcher2004},
where the column density is in $\rm cm^{-2}$ and inferred total magnetic field is in $\rm \mu G$.

All sink patches are magnetically supercritical, spanning a range $ 1 \lesssim \mu_{3D} \lesssim 8 $, and lie
in the region of magnetic supercriticality for $\log N_{H_2} \gtrsim 21.5 $.  In contrast, the``observationally'' determined magnetizations are $1  < \mu_{2D} < 4$ (Figure \ref{fig:crutcherobs}b). Thus, $\mu_{2D}$ can underpredict the values of $\mu_{3D}$ in our sample cloud cores by factors of $2-3$. A small part of the discrepancy may come from the inferred cloud geometry \citep[taken from][]{MS1976}; the rest probably lies in underestimates of the mass of the patch from the average line of sight column density, or in the differences between using the average total magnetic field $B_{TOT}$ and decomposing directional components to calculate the flux $\Phi$.

\begin{figure*}[htb]
    \centering
    \includegraphics[width=0.95\textwidth]{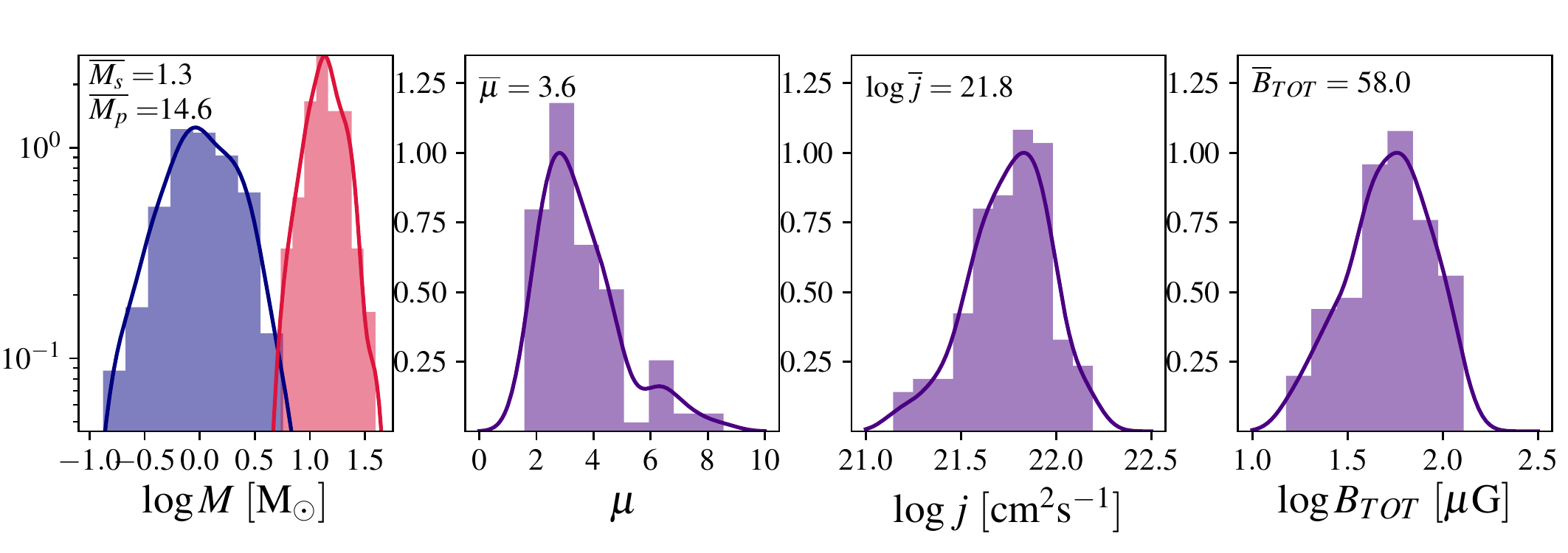}
    \caption{Distributions of core properties from left to right: mass functions for the sink mass, $M_s$, (blue) and patch gas mass $M_p$ (red), mass to flux ratio relative to critical, $\mu$, specific angular momenta, $\log j$, and total average magnetic field $\log B_{TOT}$. A total of 112 cores stacked from all the MHD runs 50,000 years after their sink formation time in the simulation were used in the sample.}
    \label{fig:coreprops}
\end{figure*}

\subsubsection{Core Angular Momenta and Magnetic Fields}
\label{sect:cores} 

\begin{figure*}[htb]
    \centering
    \includegraphics[width=0.7\textwidth]{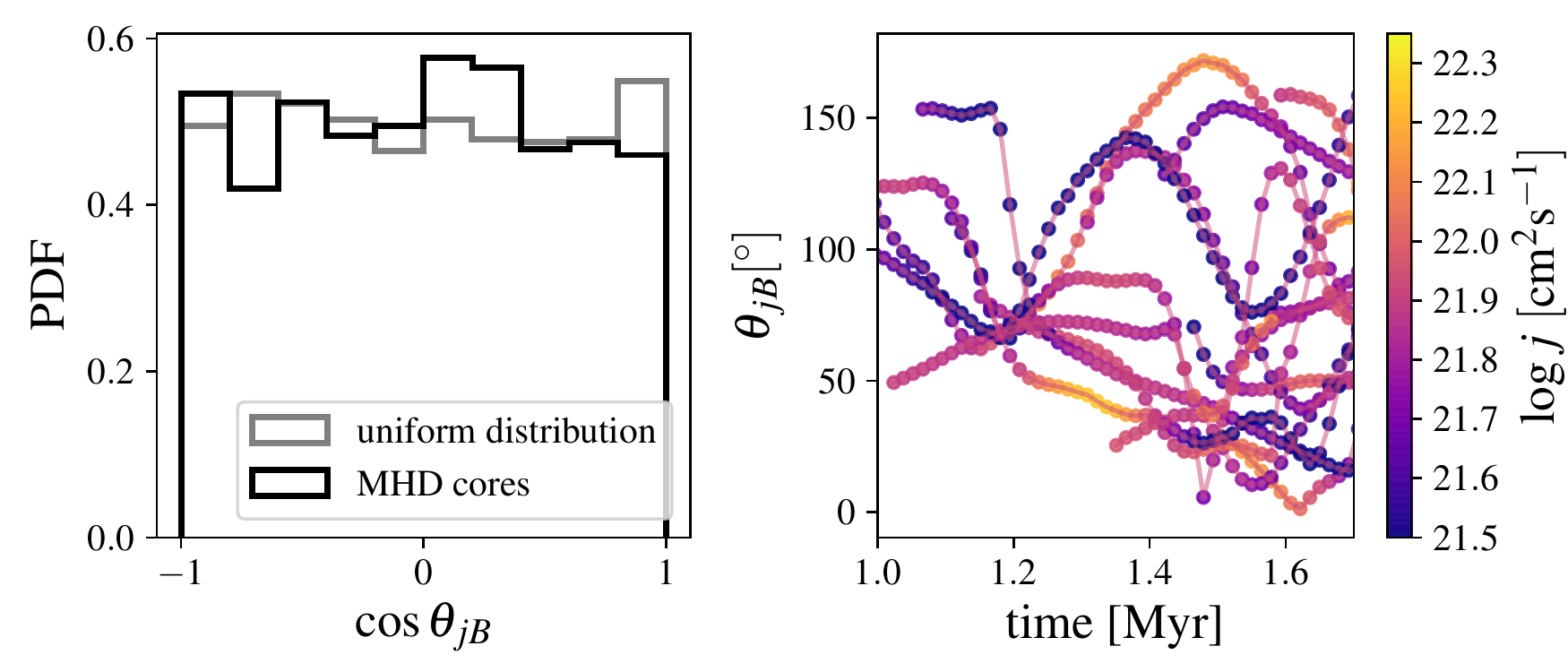}
    \caption{Behavior of the relative angle between the specific angular momentum vector and magnetic field direction $\theta_{jB}$. (Top) Distribution of $\cos \theta_{jB}$ for all cores throughout the simulations compared to a random uniform distribution. (Bottom) Sample subset $\theta_{jB}$ tracks over time from run IR\_by5\_s2, where variations are due to the changing direction of the $j$ vector, thereby also changing the magnitude of the specific angular momentum.}
    \label{fig:thetabetween}
\end{figure*}

\par To statistically characterize core properties at the $0.1 \ \mathrm{pc}$ scale, we construct a sample of sink-patch systems from every MHD run, all taken at the same respective core `age' of 50,000 years, the time elapsed since each sink's respective formation time.  All quantities, unless otherwise stated, are integrated across the entire sink-patch. We show the distributions of sink and patch gas masses, their magnetizations, the specific angular momenta, average total magnetic fields, and their probability distribution functions computed from Gaussian kernel-density estimates in Figure \ref{fig:coreprops}. The kernel-density estimates are calculated using the \texttt{scipy.stats} Gaussian kde method and the resulting probability density functions are available in the supporting data.

\par Average core angular momenta are  $j = 6.3 \pm 1.0 \times 10^{21} \ \mathrm{cm^2 s^{-1}}$ for the magnetic runs vs $j = 5.3 \pm 0.8 \times 10^{21} \ \mathrm{cm^2 s^{-1}}$ in the hydrodynamic case. As in the hydrodynamic case, the time evolution of average core specific angular momenta is flat and spin axis distributions are consistent with an isotropic distribution. 
The directions of core magnetic fields do not appear isotropic, and tend to be perpendicular to the filament orientation, as is the case for the larger scale fields (Figure \ref{fig:HRO}).  This alignment
suggests that the core fields  inherit much of their orientation from larger scales.

\par { As in the purely hydrodynamic (HD) case, cores are just as likely to be spun up or down after accretion events, leading to a flat specific angular momentum over time (see Figure 3 in \citet{kuznetsova2019} for more details)}. The magnitude of the specific angular momentum over time is highly variable over short time scales in the purely HD case; this variability is slightly dampened in the MHD case. Short-term variability in the magnitude of the specific angular momentum is correlated with changes in the direction of the angular momentum vector, which peak around  $ d \theta_{j}/dt \sim 5-10^{\circ}/10^4 \ \mathrm{yr}$. In the MHD runs, large directional changes ($d \theta_{j}/dt > 15^{\circ}/10^4 \ \mathrm{yr}$) occur at a rate of 6\%, compared to an occurrence rate of 9\% for the hydrodynamic case.

\par In comparison, fluctuations in the total magnetic field direction averaged across the sink-patch system, $B_{TOT}$, are small, on the order of $d\theta_{B}/dt \sim 1-2^{\circ}/10^{4} \ \mathrm{yr}$, and rarely exceeding $15^{\circ}/10^4 \mathrm{yr}$. Changes in the magnitude of the total magnetic field occur over longer timescales, but rarely exceed variation by more than a factor of 2 over the lifetime of the sink-patch systems.

\par As such, the changes in the angle between the spin axis of the core and that of the magnetic field, $\theta_{jB}$ are due to the variations in the direction of the core angular momentum. Misalignment between the spin axes and magnetic field direction is a natural consequence of the rapidly varying direction of angular momentum flux, leading even originally aligned cores to become misaligned over time. In Figure \ref{fig:thetabetween}, we show the distribution of $\theta_{jB}$ across all the MHD runs, the relative angle between the direction of the core magnetic field and angular momentum vectors, compared to a random uniform distribution of the same sample size. A K-S test comparison confirms that the null hypothesis can not be rejected for $\theta_{jB}$ being drawn from a random uniform distribution at 99.5\% confidence.

\subsubsection{Accretion onto Cores}

\begin{figure*}
    \centering
    \includegraphics[width=0.95\textwidth]{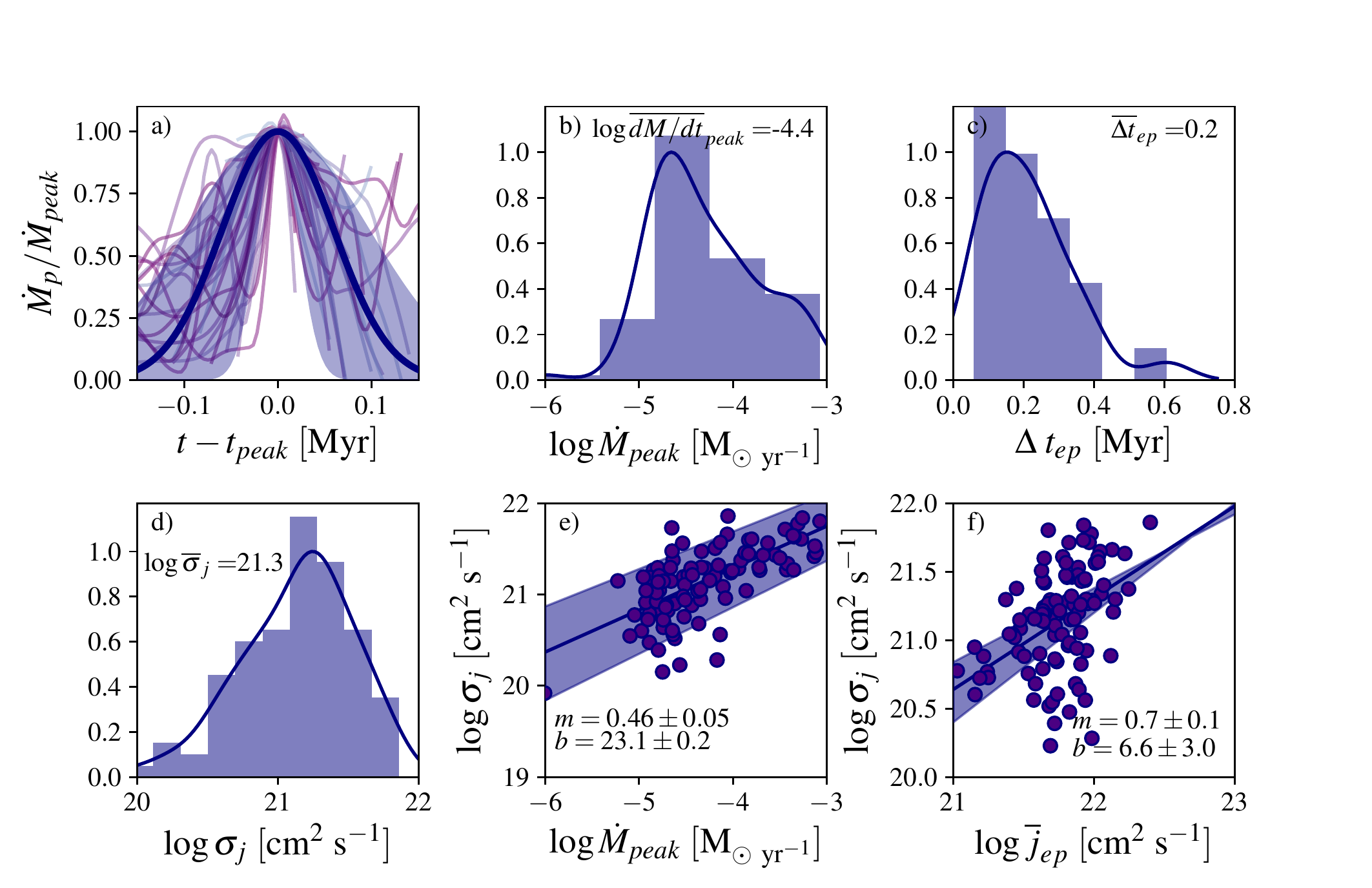}
    \caption{Episode properties for cores that had more than one episode as classified by their total net patch flux into the patch. a) Episode shapes overplotted with a Gaussian with a FWHM of 0.14 Myr. Shaded regions reflect the standard deviation in the FWHM (0.09 Myr) for all episodes b) Distribution of peak fluxes $\dot{M}_{peak}$. c) Distribution of upper limits on the total length of the episode $\Delta t_{ep}$ d) Distribution of variation of the episode angular momentum $\sigma_j$ e) Direct correlation between the variation in the angular momentum over the course of the episode and the net peak flux, with fits to $\log \sigma_j = m (\log \dot{M}_{peak} ) + b$ shown f) Correlation between the variation in the angular momentum and the mean angular momentum for the episode with fits to $\log \sigma_j = m(\log \overline{j}_{ep})+ b$. }
    \label{fig:coreacc}
\end{figure*}

\par  Variability in the flux of material accreted onto the sink-patch system plays a large role in dictating the changes in the direction of $j$ and increases the likelihood of misalignment between $j$ and $B$. Due to the heterogeneous nature of the sink-patch environment, accretion onto the core occurs episodically and non-isotropically, quite similarly to the behavior seen in the hydrodynamic case; see Figure 7 in \citet{kuznetsova2019}). Using the sample of core accretion histories and the information within the core's boundary cells we calculate the flux of quantities into the patch across the faces of patch walls. Analyzing the accretion over time, we statistically characterize patch accretion episodes by their length, peak flux, and effect on the specific angular momentum.

\par For each core in the sample of MHD runs, we identify accretion episodes first by finding the roots of a polynomial fit to the total net accretion onto the patch over time, $\dot{M}_p$; the start and end times for each episode are then given by the location of the polynomial's zeroes.
For each episode we calculate a peak flux amount, an episode length, and a variation in the specific angular momentum, $\sigma_j$. The peak flux amount $\dot{M}_{p,peak}$ is the maximum of the total net flux into the patch over the duration of the episode $\Delta t_{ep}$. The standard deviation of the angular momentum on the boundary during the episode is then $\sigma_j$, compared to the mean episode angular momentum $\overline{j}_{ep}$. 

\par In Figure \ref{fig:coreacc}, we show the accretion properties across episodes for every core that had more than one accretion episode, as identified by the behavior of the total flux into the patch. The total net flux into the patch leads to a fairly conservative classification scheme that excludes the cores that appear to experience continuous accretion. However, even when the total net flux appears continuous, decomposing the total flux into orthogonal components yields many smaller overlapping sub-episodes as the cores switch between spatially distinct accretion reservoirs (see Appendix \ref{sect:subep} for details). The overlap between directional components in time makes it difficult to accurately assign properties to individual sub-episodes, thus we only consider the total net flux when discussing properties of the episodes. This means that the episode lengths in Figure \ref{fig:coreacc}c are upper limits; component sub-episodes tend to be a factor of 2-3 shorter in duration.

\par Normalized by the patch flux, and centered on when the patch flux is at a maximum, the average episode shape can be modeled by a Gaussian centered on the time of the episode's peak accretion with $\sigma=0.06$, for a FWHM of $0.14 \rm Myr$.  A typical peak mass flux into the patch is $ 3.6 \times 10^{-5} \ \mathrm{M_{\odot} yr^{-1}}$ at $0.1 \ \mathrm{pc}$. As shown in Figure \ref{fig:coreacc}e-f, the peak mass flux into the patch and the mean specific angular momentum for an episode are correlated with the magnitude of the variation in the angular momentum during the episode. For a typical core accretion episode, the expected variation in the specific angular momentum is on the order of $\log \sigma_j = 0.2 \log  \overline{j}_{ep}$. 

\par Integrating the mass accreted in an episode, and comparing contributions from each orthogonal direction, we can place a lower limit on the frequency of anisotropic accretion.  We find that at least $\sim 60\%$
of episodes as we've characterized them above have accreted primarily along only one or two directions. This means that for the majority of episodes, accretion occurs in a filament or sheet-like flow.

\section{Discussion}

\subsection{Global Magnetic Field Behavior}

It is important to test whether our simulations are in reasonable agreement with observations of magnetic field strengths and structures on scales larger than cores. As discussed in \S \ref{sec:globalb}, the variation of magnetic field strength with density is consistent with the observational results of \cite{crutcher2010} (Figure \ref{fig:crutcher}). As discussed in \citet{crutcher2010,crutcher2012}, this behavior is consistent with flow along field lines at low densities,  and the scaling law predicted from  spherical gravitational contraction during flux freezing, $B \propto \rho^{2/3}$, at high densities.  Based on Figure \ref{fig:crutcher}, flux-freezing is maintained up to the highest densities in the simulation, constituting an important check of our simulations with observational results. The orientations of the plane of sky magnetic fields relative to filamentary structure are also consistent with observational results from \citep{planck2016} (Figure \ref{fig:HRO}). In the Appendix, we show maps of the angle $\phi$ between the magnetic field ${\bf B}$ and the iso-column density contours, which at least qualitatively show the same jumbled structure seen in Figure 9 of \citet{planck2016}.

\subsection{Magnetic Flux, Misalignment, and the Magnetic Braking Catastrophe}

\par Investigations of the effects of misalignment between the magnetic field and spin axis on disk formation, found that both misalignment on the order of $20^{\circ}$ and magnetic supercriticality of at least $\mu \geq 4$ are required to bypass or reduce the effects of the magnetic braking catastrophe without invoking non-ideal MHD effects \citep{mellon2008}. In this study we find initial mass to flux ratios with a mean around $\mu = 3.6$, in a distribution with fairly long tail that extends out to values of $\mu \sim 10$ (Figure \ref{fig:coreprops}b).  In this context it is important that our volumetrically-derived magnetizations $\mu_{3D}$ are systematically larger than the $\mu_{2D}$ calculated using the \citet{crutcher2004} relation (Figure \ref{fig:crutcherobs}a).

\par We note also that the value of $\mu$ increases over time as cores accrete mass more readily than their magnetic field values increase - a result which follows from accretion along field lines. Misalignment between the magnetic field and the spin axis of the core is common; at the core scale ($r \sim 0.1 \ \mathrm pc$), we find a random uniform distribution of angles between the magnetic field vector and the spin axis of the core (Figure \ref{fig:thetabetween})a.\citet{krumholz2013} concluded that the observed protostellar disk fraction can not be reproduced with a random uniform distribution of $\theta_{jB}$ and a median mass to flux ratio of 2, based on the observed mass to flux from the surveys of molecular cloud cores by \citet{crutcher2012}. This problem could be at least partially addressed with the larger mass to flux ratios we've calculated for the cores in this study. However, even if we were to discount the difference between $\mu_{2D}$ and $\mu_{3D}$, their analysis does not account for a time variable $\theta_{jB}$ in which any core's spin axis has a chance of becoming misaligned relative to the magnetic field at some point in its evolution, a process we show to be quite common in our simulations in Figure \ref{fig:thetabetween}b. 

\par Due to the time variability in the direction of the angular momentum,  alignment at any one time does not guarantee alignment at later times, such that any core can become misaligned from the magnetic field after an accretion episode. Whereas, \citet{joos2013} produces misalignments by imposing a turbulent velocity within the core, we find that misalignment can be externally imposed by episodic highly directional accretion flows from a local heterogeneous environment. Recent simulations from \citep{wurster2019} of star formation in turbulent magnetized clumps find that protostellar disk formation occurs independent of the presence of non-ideal MHD effects  and conclude that turbulent gas motions of the protostellar core environment play a larger role than magnetic effects in the formation of disks.

\par As a caveat, due to the resolution of the sink-patch, $\theta_{jB}$ is effectively a measurement on the $0.1 \ \mathrm{pc}$ scale, while disk formation simulations induce misalignment extending down to the protostar. It is possible that significant angular momentum transport on sub-core scales could dampen the effects seen at our resolution limit.

\subsection{Hallmarks of a Heterogeneous Star Forming Environment}

\par  On large scales, filaments in the simulation are in good agreement with observed star forming regions when smoothed to similar resolutions, but it is clear that the inter-filament medium is highly substructured. The episodic nature of accretion captured at the core scale implies a heterogeneous star-forming environment, a fact also noted in other numerical studies of cluster formation \citep{kuffmeier2017,kuffmeier2018}. The time dependent directionality of accretion onto cores creates the variation in the direction of the angular momentum vector and suggests the existence of spatially distinct accretion reservoirs. That is, there must be a finite number of overdensities in the form of directional accretion flows or streams in order to explain the amount of variation in the direction of the angular momentum vector.

\par Filament `fibers' have been identified by the presence of velocity coherent substructures within observations of star-forming regions \citep[e.g.][]{Hacar_2013,hacar2018}. While there are many proposed explanations for this sub-structuring, including fragmentation, it is possible that these sub-filamentary structures are observational signatures of the type of dense structures that fall in to produce episodic core accretion and changes in the direction of the angular momentum vector. To support this conjecture,
a fiber of length $\sim 0.1$~pc, characteristic of structures seen in our simulations and onsistent with the peak of the ISF fiber lengths in \citet{hacar2018}, next to a core of
mass $10-15 \ \rm M_{\odot}$ would exhibit an infall velocity of $0.7-0.8 \rm \ km \ s^{-1}$, resulting in an accretion episode lasting $\sim 0.14$ Myr, consistent with the simulations (upper right panel of Figure \ref{fig:coreacc}). Note that for typical magnetic field values of $60 \ \rm \mu G$ at densities of at least $n_{H} \sim 10^5 \rm cm^{-3}$, the Alfven velocity is at most $0.3 \ \rm km \ s^{-1}$, and so the inflow is at least marginally super-Alfvenic. 

\subsection{Considerations for Modeling Realistic Disk Formation}

\par We find that the ideal MHD model used here decreases the efficiency of star formation compared to our hydrodynamic runs, while the cores that do form are supercritical such that their formation and evolution is dominated by gravitational infall. The statistical relationship between the patch masses and sink accretion rates such that $\dot{M}_s = (M_s+M_p)^2$ investigated in \citep{Kuznetsova_2018b} remains true for the cores in this sample. The presence of magnetic fields does not impact the growth of the upper-mass power law by accretion;  gravitational infall drives accretion onto the sink-patch system.

\par The dynamics of the cores are primarily dominated by the influx of multidirectional flows, as in \citep{kuznetsova2019}. The shapes of the distributions and behavior of the specific angular momenta between the HD and MHD runs are also consistent. We do find that the MHD runs exhibit lower amplitude variability in the magnitude of the specific angular momentum and experience fewer intense directional flips than the HD case. This suggests that the magnetic field may have some influence on the direction of the angular momentum vector, but not enough to enforce alignment, at least on the core scale. 

\par The specific angular momentum remains nearly constant over time for all cases, showing essentially the same behavior as in the hydrodynamic case (see Figure 3 in \citet{kuznetsova2019}). This behavior suggests that growing the core through accretion from a realistic heterogeneous environment has marked departures from the disks predicted by the smooth isotropic disk formation models which predict growth of specific angular momentum with time as angular momentum is inherited from a uniformly rotating cloud \citep{TSC_84}. 

\par Protostellar collapse models have become increasingly more sophisticated as they include additional physics such as non-ideal MHD or radiation effects \citep[e.g.][]{wurster2018,bhandare2018}, but the initial conditions adopted in these studies are commonly smooth and spherically symmetric. Other top-down numerical studies which simulate the dynamics of the molecular cloud as well as zoom in to the formation of individual protostars and disks also find that accretion is very much non-isotropic, commonly occurring in filaments or sheets and time variable in nature \citep{kuffmeier2017}. While additional transport processes occur on scales that are non-resolvable in our simulations, the statistical assemblages presented here offer a range and distribution of initial and boundary conditions for future work with more detailed disk simulations.
\par We provide the probability density functions and of both core and infall properties and requisite code tools as a supplementary archive \footnote{https://github.com/akuznetsova/proto-props}. 

\section{Summary} 

\par We modeled the formation of star clusters with the grid code \emph{Athena} including ideal MHD effects in order to understand how magnetic fields affect star formation on both the cloud and core scales. Our main findings were:
\begin{enumerate}
    \item Globally, simulations were in agreement with observed magnetic field structures measured on the same scales.
    \item A comparison of `observed' and actual core magnetizations yields that is possible to underpredict $\mu$ from projected data, with a factor of $\sim 2 $ difference in the median $\mu$, compared to reported values from Zeeman measurements of molecular clouds.
    
    \item The mass inflows into protostellar cores have specific angular momenta $j$ that are randomly oriented with respect to the overall magnetic field direction.  The inflows often occur in distinct episodes lasting $\gtrsim 0.1 $ Myr, with each individual episode exhibiting very different orientations of $j$. There is little time evolution of the 
magnitude of $j$.  
\end{enumerate}

\par  Our findings indicate that protostellar accretion is likely not characterized by slowly varying, isotropic infall often assumed. The distributions of core properties, as well as characterisation of discrete episodes of accretion onto the cores presented in this paper, can provide more realistic initial and boundary conditions for studies of protostar and disk formation in higher resolution simulations.

\acknowledgements
This work was supported by NASA grant NNX16AB46G and by the University of Michigan, and used computational resources and services provided by Advanced Research Computing at the University of Michigan, Ann Arbor and the University of North Carolina at Chapel Hill Research Computing group.


\appendix
\renewcommand{\thefigure}{A\arabic{figure}}
\setcounter{figure}{0}

\section{Large Scale Magnetic Field Structure}

\subsection{Calculating relative orientations}
\label{sect:hro}
\par  The relative orientation is computed as the angle $\phi$, the angle between the magnetic field ${\bf B}$ and the iso-column contours, is represented by the direction of the gradient in column density $N_H$. The gradient is calculated by convolving the column density map  with  a symmetric two-dimensional Gaussian  derivative  kernel, $G_x$ and $G_y$ are calculated using the $x$- and $y$- derivatives, respectively:

\begin{equation}
\nabla N_H = (G_x \otimes N_H) {\bf \hat{i}} + (G_y \otimes N_H) {\bf \hat{j}} = g_x  {\bf \hat{i}} + g_y {\bf \hat{j}}
\end{equation}

\begin{figure}[h!]
    \centering
    \gridline{\fig{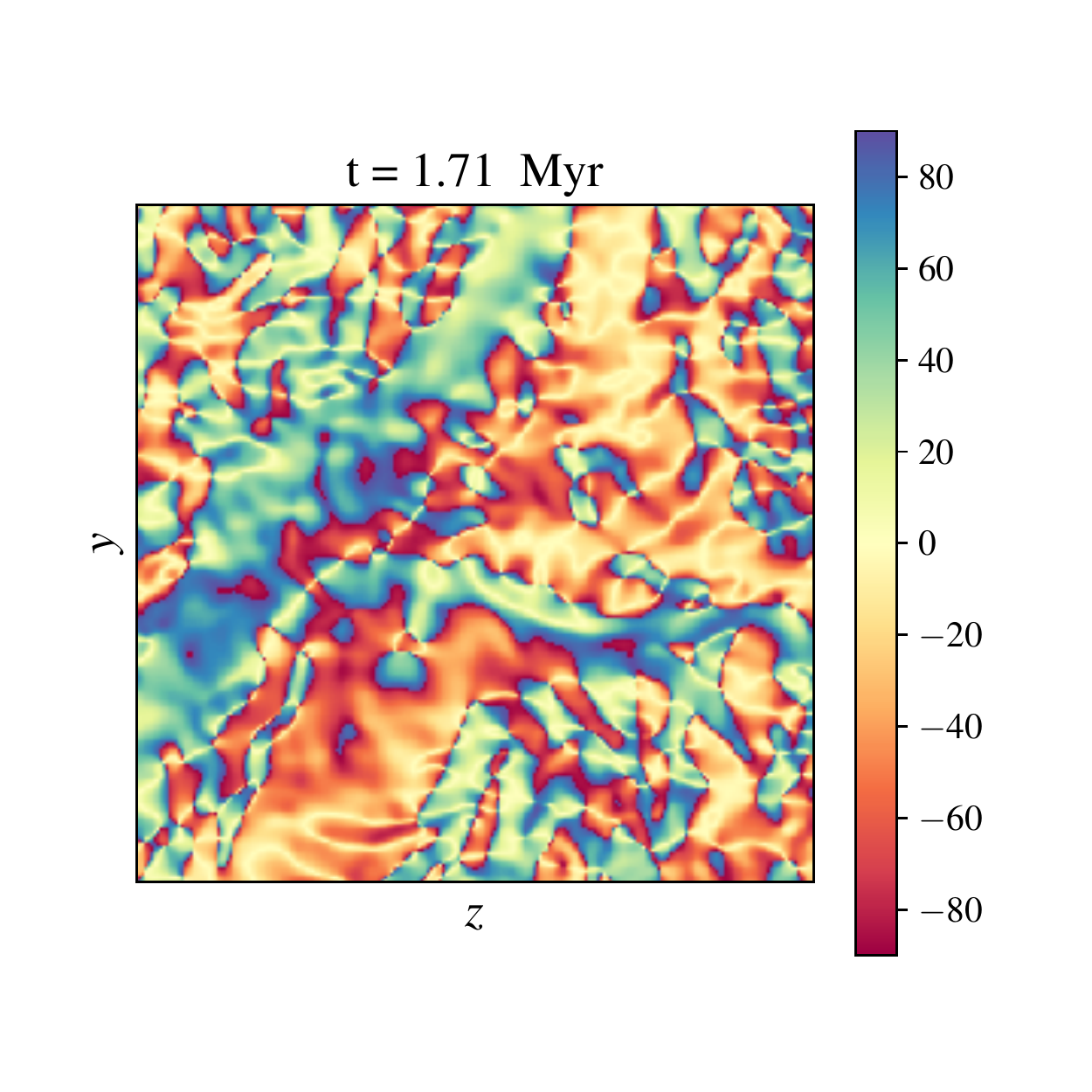}{0.33\columnwidth}{} \fig{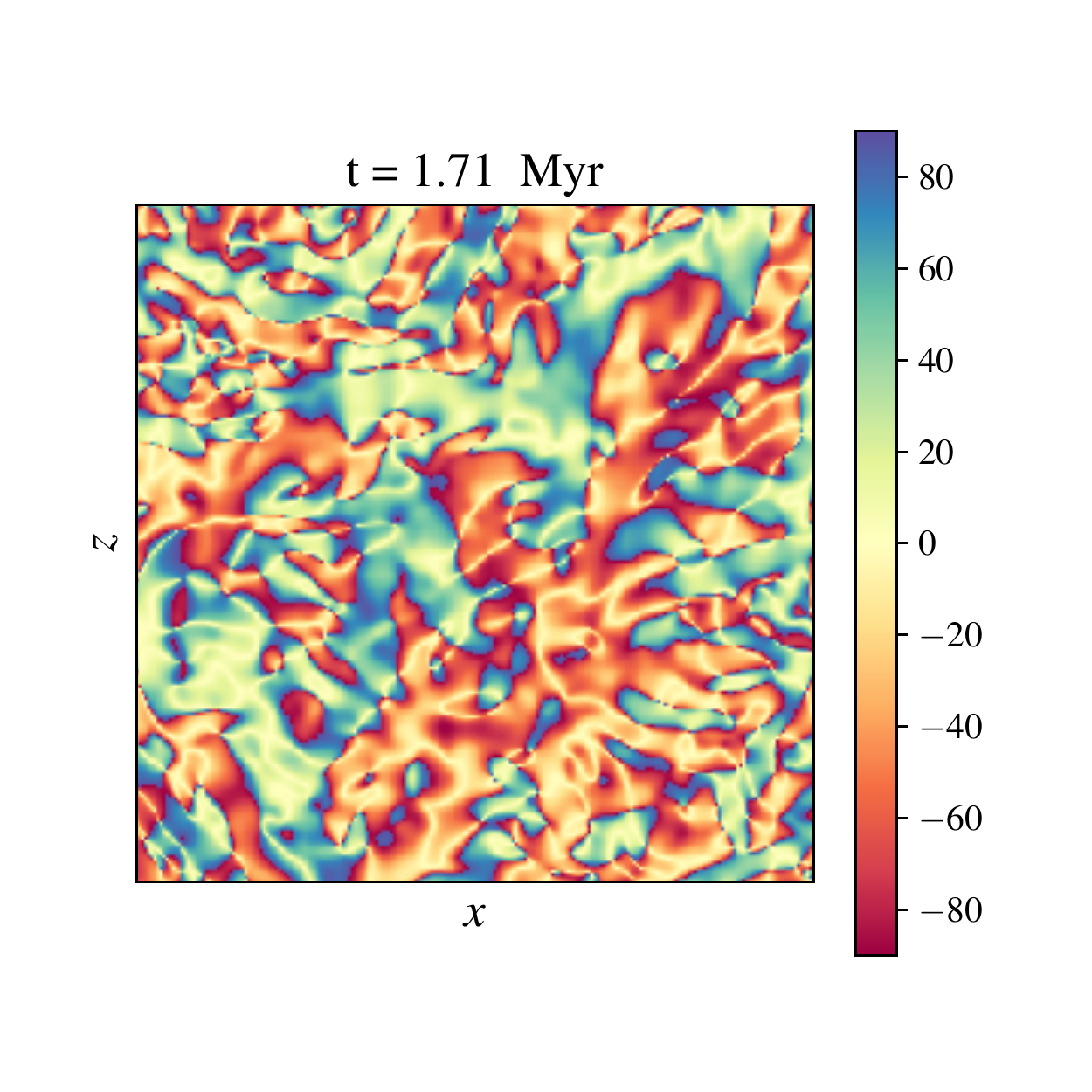}{0.33\columnwidth}{} \fig{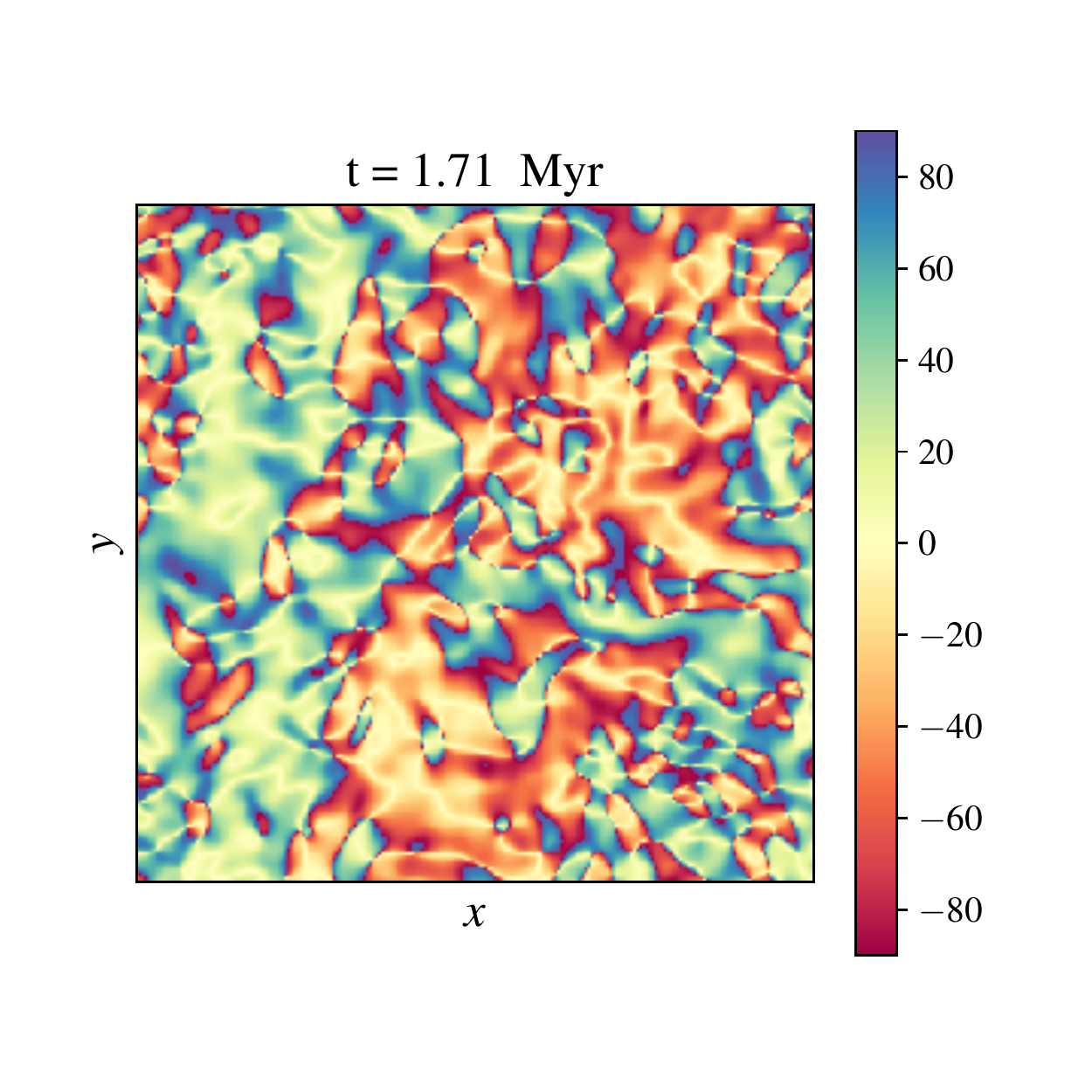}{0.33\columnwidth}{}}
    \caption{Maps of the angle between the the magnetic field direction and the iso-density contour, $\phi$, for column density projections of run IR\_by5\_s2 across the inner 12 pc of the simulation domain, centered on the cloud.}
    \label{fig:phimap}
\end{figure}

The orientation of the gradient is then $\theta_{N_H} = \arctan(-g_x, g_y)$. We show the maps of $\phi$ for all three projections in Figure \ref{fig:phimap}. The results show reasonable agreement with observations \citep[see Figure 9 in][]{planck2016}.

\par Based on the methods outlined in \citet{planck2016}, we construct a histogram of relative orientation (HRO) between the derived filament and the magnetic field directions. Relative orientations are measured as the angle between the magnetic field in each cell and the gradient across the column density for each projection of run IR\_by5\_s2. Both the magnetic field and column densities are smoothed with a Gaussian kernel of $\sigma = 4$ cells, to limit to resolved structures on the $0.4 \mathrm{pc}$ scale. This chosen smoothing corresponds to the smoothed Planck map resolution of 10' at a distance of $140 \mathrm{pc}$, similar to the Taurus region. 

\par Results are evenly sampled from column density ranges at the 68th, 90th, and 98th percentile level, each corresponding to a line in Figure \ref{fig:HRO}. At the chosen 90th percentile level, $\log N_H/cm^{-2} = 21.7$, the HRO becomes very clearly in favor of a perpendicular orientation between the magnetic field and the iso-density contour; this value is consistent with the turnover density determined from the Planck polarisation measurements. The errors due to binning shown as a shaded region and calculated from $\sigma_h^2 = N_i(1-N_i/N_{tot})$ , where $N_i$ and $N_{tot}$ are the number of samples in the bin and the the number of total samples, respectively. 

\par  In Figure \ref{fig:HRO}a-b, the magnetic fields are preferentially oriented perpendicularly to the highest density filaments, with some parallel alignment in the lower density regime. The parallel alignment of the lowest density contours is difficult to discern by eye in the HRO, but the relative integrated contributions from the center of the HRO are greater compared to those from the outer edges. In the case of the projection in the $x-y$ plane, where the line of sight is preferentially along the filament axis, the orientation is marginally in favor of parallel orientations, rather than perpendicular ones.

\section{Episodes and Sub-episodes}
\label{sect:subep}
\par
\begin{figure*}
    \centering
    \includegraphics[width=0.9\textwidth]{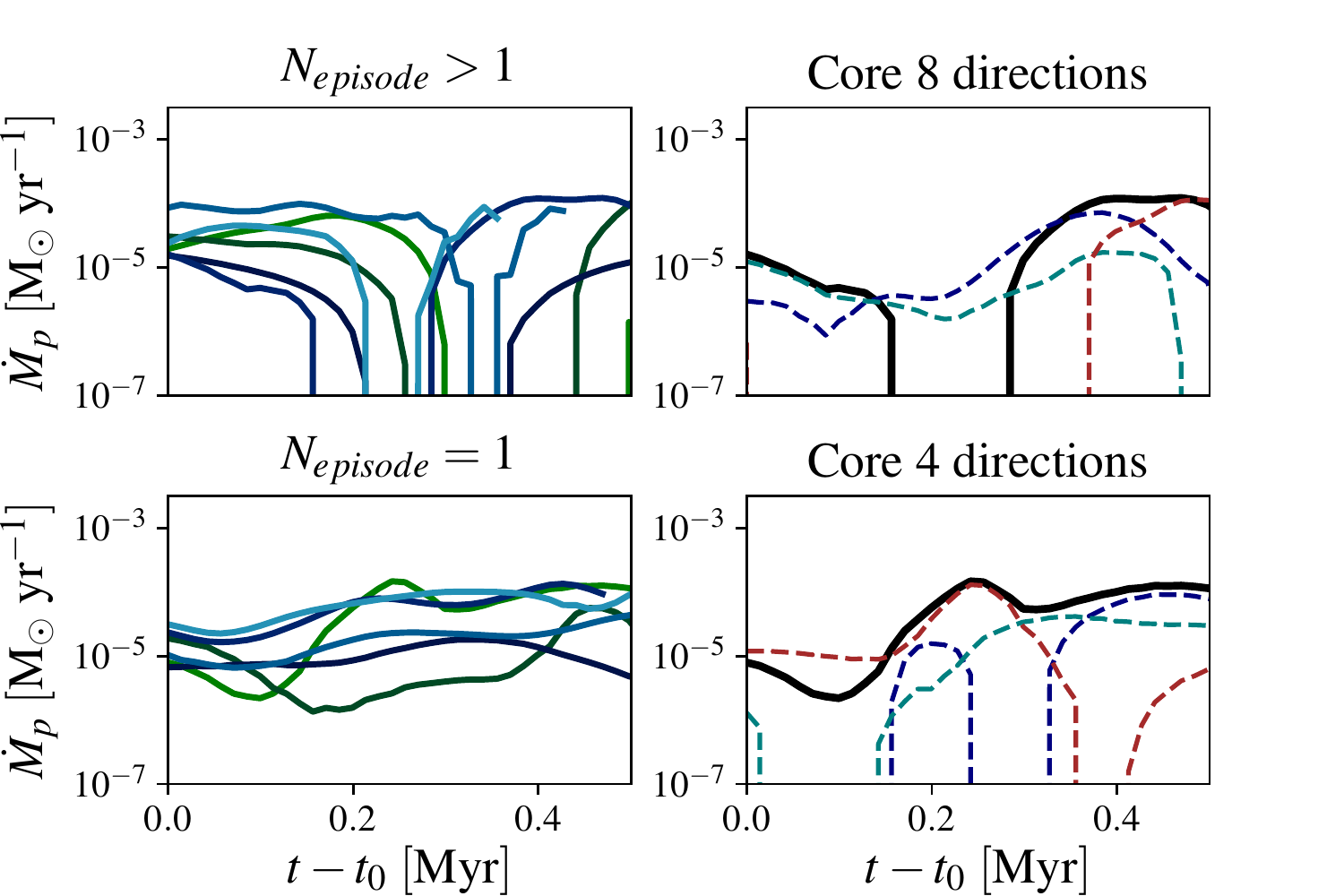}
    \caption{Top row: Patch fluxes for cores with more than one episode (left) sample of cores with more than one episode (right) core 8's patch flux (solid black line) decomposed into its orthogonal components (dashed lines). Bottom row: Patch fluxes whose total flux is continuous (left) sample of cores with continuous total patch flux (right) continuous patch flux for core 4 (solid black line) with its component parts, decomposed into orthogonal directions (dashed lines). }
    \label{fig:subep}
\end{figure*}

 The sample of core episodes only includes cores that have more than one episode as classified by the total net flux into the patch. This classification removes cores that appear to be accreting continuously. However, cores with continuous total patch accretion over time are not necessarily isotropically and smoothly accreting from their surrounding medium (see core 4 in Figure \ref{fig:subep}). When the total net flux is decomposed into its orthogonal components, it is evident that the larger continuous episode is comprised of multiple periods of time when different flows dominate, suggesting accretion from discrete spatial sources. 
\par The fact that episodes in total patch flux are actually comprised of many sub-episodes also applies to the sample of cores used in the study, those with $N_{episode} > 1$, such as core 8's last episode seen in Figure \ref{fig:subep} in which a transition from one direction to another happens halfway through. This overlap makes it difficult to constrain a sub-episode's contribution to the mass accreted, for example, as some directions are also flows out of the patch, which is why we have opted to use the total net flux in our analysis. However, we note that this means that the distributions of episode lengths are then upper limits. 

\par 

\bibliographystyle{aasjournal}
\bibliography{biblio}
\end{document}